\documentclass[journal,final]{IEEEtran}

\usepackage{amssymb}

\usepackage[utf8]{inputenc}
\usepackage{tipa}
\usepackage{relsize}

\usepackage[english]{babel}
\usepackage[autostyle]{csquotes}

\usepackage{amssymb,amsmath}

\usepackage[detect-all,binary-units,tight-spacing]{siunitx}
\usepackage{graphicx}

\usepackage{tabulary}
\usepackage{booktabs}

\usepackage{caption}
\DeclareCaptionLabelSeparator{periodquad}{.\quad}
\usepackage[inline,shortlabels]{enumitem}

\usepackage[obeyFinal,colorinlistoftodos]{todonotes}
\usepackage{acro}
\DeclareAcronym{2D}{long=two-dimensional}
\DeclareAcronym{3D}{long=three-dimensional}
\DeclareAcronym{BAP}{long=aperiodicity per band}
\DeclareAcronym{CAPT}{long=computer-assisted pronunciation training}
\DeclareAcronym{DFKI}{long=German Research Center for Artificial Intelligence}
\DeclareAcronym{DoF}{short-plural=, long=degree of freedom, long-plural-form=degrees of freedom}
\DeclareAcronym{EGG}{long=electroglottography}
\DeclareAcronym{EMA}{long=electromagnetic articulography}
\DeclareAcronym{EPG}{long=electropalatography}
\DeclareAcronym{EST}{long=Edinburgh Speech Tools}
\DeclareAcronym{F0}{short={F\textsubscript{0}}, long=fundamental frequency}
\DeclareAcronym{GMM}{long=Gaussian mixture model}
\DeclareAcronym{HMM}{long=hidden Markov model}
\DeclareAcronym{HTS}{long=\ac{HMM} based synthesis}
\DeclareAcronym{MCD}{long=mel cepstral distortion}
\DeclareAcronym{MGC}{long=mel-generalized cepstral coefficients}
\DeclareAcronym{MLSA}{long=mel log spectrum approximation}
\DeclareAcronym{MRI}{long=magnetic resonance imaging}
\DeclareAcronym{PCA}{long=principal component analysis}
\DeclareAcronym{RMSE}{long=root mean square error}
\DeclareAcronym{TTS}{long=text-to-speech}
\DeclareAcronym{UTI}{long=ultrasound tongue imaging}
\DeclareAcronym{VUV}{long=voiced-unvoiced}
\DeclareAcronym{XRMB}{long=X-ray microbeam}
\DeclareAcronym{HOSVD}{long=higher order singular value decomposition}

\usepackage[hidelinks]{hyperref}
\usepackage[all]{hypcap}
\usepackage[capitalise, nameinlink]{cleveref}
\crefname{equation}{}{}

\usepackage[natbib, style=ieee, maxbibnames=999, maxcitenames=2, mincitenames=1]{biblatex}
\begin{document}

\setlength{\tabcolsep}{0.1em}

\title{Synthesis of Tongue Motion and Acoustics From \\
  Text using a Multimodal Articulatory Database}

\author{Ingmar~Steiner,
        Sébastien~Le~Maguer,
        and~Alexander~Hewer%
\thanks{All authors are with the Cluster of Excellence \enquote{Multimodal Computing and Interaction,} Saarland University, Saarbrücken 66123, Germany
  (e-mail: \url{steiner@coli.uni-saarland.de}; \url{slemaguer@coli.uni-saarland.de}; \url{hewer@coli.uni-saarland.de}).
  
  This paper has supplementary downloadable material available at \url{http://ieeexplore.ieee.org}.
  This includes videos of all synthesized utterances in the test set, produced using the tongue model parameters (cf., \cref{sec:weights}).
  This material is \SI{39}{\mega\byte} in size.
  
  Digital Object Identifier \href{http://dx.doi.org/10.1109/TASLP.2017.2756818}{10.1109/TASLP.2017.2756818}}%
}

\markboth{IEEE/ACM Transactions on Audio, Speech, and Language Processing}%
{Author \MakeLowercase{\textit{et al.}}: Submission}

\maketitle

\begin{abstract}
We present an end-to-end text-to-speech (TTS) synthesis system that generates audio and synchronized tongue motion directly from text.
This is achieved by adapting a 3D model of the tongue surface to an articulatory dataset and training a statistical parametric speech synthesis system directly on the tongue model parameters.
We evaluate the model at every step by comparing the spatial coordinates of predicted articulatory movements against the reference data.
The results indicate a global mean Euclidean distance of less than 2.8 mm, and our approach can be adapted to add an articulatory modality to conventional TTS applications without the need for extra data.

\end{abstract}


\section{Introduction}
\label{sec:intro}

\IEEEPARstart{T}{he} sound of human speech is the direct result of production mechanisms in the human vocal tract.
Air flows from the lungs through the glottis, whose vocal folds can be set to vibrate, the sound of which is then filtered by the shape of the tongue, lips, and other articulators, generating what we perceive as audible signals such as spoken language.
Researchers in phonetics and linguistics have studied these speech production mechanisms for many years, but while the acoustic signal and facial movements can be observed and measured directly, doing the same for partially or fully hidden articulators such as the tongue and glottis is not as straightforward.

Consequently, sensing and imaging techniques have been applied to the challenge of observing speech production mechanisms \emph{in vivo}, which has greatly improved our understanding of these processes.
The corresponding modalities include, fluoroscopy \cite{Munhall1995JASA}, \ac{UTI} \cite{Stone2005}, \ac{XRMB} \cite{Westbury1994}, \ac{EMA} \cite{Schoenle1987, Hoole2010}, and real-time \ac{MRI} \cite{Niebergall2012, Narayanan2014}, among others.
Some of these involve health hazards (due to ionizing radiation), and all are more or less invasive, but they produce \emph{biosignals} which, in combination with simultaneous acoustic recordings, represent multimodal articulatory speech data.
The benefits are tempered by the challenges of processing the imaging and/or point-tracking data, which in the field of speech processing has created new opportunities for collaboration with areas such as medical imaging and computer vision.

The biosignals that can be obtained using such modalities to record spoken language, provide opportunities to greatly enhance models of speech by integrating measurements of the underlying processes directly with the acoustic signal.
This leads to more elegant and powerful approaches to speech analysis and synthesis \cite{Ling2009, Ling2010SpeCom, Richmond2015}.
However, it must be borne in mind that all of the biosignals produced by the modalities mentioned above represent a sampling of the articulators that is \emph{sparse} in the temporal domain, the spatial domain, or both.

Depending on the manner in which the data is used for analysis or applications, the resolution may need to be increased, but the missing samples cannot be restored without prior knowledge, typically provided by a statistical model trained on other data.

In this study, we present an approach to multimodal \ac{TTS} synthesis that generates the fully animated, \ac{3D} surface of the tongue, synchronized with synthetic audio, using data from a single-speaker, articulatory corpus that includes \ac{EMA} motion capture of three tongue fleshpoints \cite{Richmond2011}.
The audio and articulatory motion are synthesized using the \ac{HTS} framework \cite{Zen2009}, while the surface restoration is performed by means of a multilinear statistical tongue model \cite{Hewer2018CSL} trained on a multi-speaker, volumetric \ac{MRI} dataset \cite{Richmond2012}.
The potential application domains of this approach include audiovisual speech synthesis and \ac{CAPT}, among others.

\subsection{Background}
\label{sec:background}

Deriving models suitable for producing speech related tongue motion is an active field of research.
Such models can, for example, help to analyze and understand articulatory data that is very sparse in the spatial domain.
Ideally, such tongue models should offer a good compromise between accuracy of the generated shape and the available \acp{DoF} for manipulating it.
This means that biomechanical models such as those presented by \citet{Lloyd2012}, \citet{Xu2015}, \citet{Wrench2015}, or \citet{Yu2017} might be too complex for this purpose.
While such models aim to simulate the underlying mechanics of the human tongue as closely as possible, and can be used to visualize existing articulatory data, they can be challenging to control efficiently.

Geometric tongue models are less complex than their biomechanical counterparts.
Here, we distinguish between generic and statistical tongue models.
Generic tongue models are 3D models of the tongue that may be deformed and animated by using standard methods in computer graphics.

Statistical tongue models, on the other hand, are constructed by analyzing the \acp{DoF} of the tongue shape in recorded articulatory data, such as \ac{MRI} recordings of speech related vocal tract shapes.
Roughly speaking, such an analysis can be carried out in two ways.
The first variant investigates shape variations related to the tongue pose that are specific for speech production.
Examples of such approaches are the works by \citet{Engwall2000, Badin2002}, and \citet{Badin2006}, who examined those variations in \ac{3D} \ac{MRI} scans from a single speaker, respectively.
These methods only estimate the \ac{DoF} that are tongue pose related, while shape variations that may describe anatomical differences are missing.

Another class of methods aims at investigating those anatomy and tongue pose related shape variations separately.
This paradigm offers several advantages:
First, the results give access to tongue models that may be adapted to new speakers.
Second, this type of analysis may also provide insight into how anatomical differences affect human articulation.
For \ac{2D} \ac{MRI}, such work was conducted, e.g., by \citet{Hoole2000} and \citet{Ananthakrishnan2010}.
\citet{Zheng2003} investigated those variations in a sparse point cloud extracted from 3D \ac{MRI}.
Most recently, we performed such an analysis on mesh representations of the tongue that were extracted from 3D \ac{MRI} scans \citep{Hewer2018CSL}.

Such geometrical models have been successfully used in previous work to generate animations from provided articulatory data:
\citet{Katz2014} presented a real-time visual feedback system that deforms a generic tongue model using \ac{EMA} data.
However, due to the generic nature of the model, their approach did not take anatomical differences into account.
A statistical model was used in the approach by \citet{Badin2008}, who used volumetric imaging data of one speaker to derive the tongue model, and \ac{EMA} data of the same speaker to animate it.
\citet{Engwall2003} followed a similar approach.
Our own previous work utilized a multilinear statistical model to visualize \ac{EMA} data, which allowed it to be adapted to different speakers \citep{James2016}.

Independently, there is a growing body of work on application-oriented research to combine articulatory data, and features derived from it, with speech technology applications, such as to recover articulatory movements from the acoustic signal (\enquote{articulatory inversion mapping}, cf.\ \citep{King2007, Mitra2011} for examples), provide articulatory control for reactive \ac{TTS} synthesis (e.g., \citep{Astrinaki2013, Ling2013}),
or predict sparse articulatory movements from a symbolic representation (e.g., \citep{Ling2010SpeCom, Cai2015}).

Early studies on animating full 3D tongue surface models using \ac{EMA} data for multimodal speech synthesis, such as those of \citet{Engwall2002ICSLP} or \citet{Fagel2004SpeCom}, used concatenative \ac{TTS} systems.
Other approaches (e.g., \citep{BenYoussef2011PhD}) for \ac{HMM} based \ac{TTS} with intra-oral animation also rely on acoustic-articulatory inversion mapping.
However, to our knowledge, no previous study has presented an end-to-end system to directly synthesize acoustics and the motion of a full \ac{3D} model of the tongue surface from text using statistical parametric speech synthesis, particularly with a tongue model that can be easily adapted to the anatomy of different speakers.

\section{Method}
\label{sec:method}

\subsection{Multilinear Shape Space Model}
\label{sec:tonguemodel}

In our approach, we utilize a multilinear model to describe different tongue shapes.
This is achieved by using this model to create a function

\begin{equation}
  f : \mathbb{R}^m \times \mathbb{R}^n \to \mathcal{M}
\end{equation}

that maps the parameters \( \vec{s} \in \mathbb{R}^m \) and \( \vec{p} \in \mathbb{R}^n \) to a polygon mesh \( M = ( V, F ) \in \mathcal{M} \).
Such a mesh consists of a vertex set \( V := \{ \vec{v}_i \} \) that contains positional data \( \vec{v}_i \in \mathbb{R}^3 \) and a face set \( F \) that uses these vertices to form the collection of surface patches of the represented shape.
We note that these meshes \( M \) have the same face set and only differ in the positional data of their vertices.
The used parameters in the function describe two distinct sets of features:
On the one hand, the speaker parameter \( \vec{s} \) determines the anatomical features of the generated tongue.
The pose parameter \( \vec{p} \), on the other hand, represents the shape properties that are related to articulation.

To compute the multilinear model, we use a database that consists of \ac{MRI} scans of \(m\) speakers showing their vocal tract configuration for \(n\) different phonemes.
By means of image processing and template matching methods, we extract tongue meshes \( M \in \mathcal{M} \) from the \ac{MRI} data, such that in the end, for each speaker, one mesh is available for each considered phoneme.
This processing is described in detail by \citet{Hewer2016PSA, Hewer2018CSL}.
We then proceed to derive the \ac{DoF} of the anatomy and speech related variations.
To this end, we center the obtained meshes and turn them into feature vectors by serializing the positional data of their vertices.
Afterwards, we construct a tensor \(A\) of third order consisting of these feature vectors, such that the first mode of the tensor corresponds to the speakers, the second one to the considered phonemes, and the third one to the positional data.

In a final step, we apply \ac{HOSVD} \cite{Tucker1966} to obtain the following tensor decomposition:
\begin{equation}
  A = C \times_1 U_1 \times_2 U_2
\end{equation}

In this decomposition, the tensor \( C \) is of third order and represents our multilinear model.
The operation \(C \times_n U\) is the \(n\)-th mode multiplication of the tensor \(C\) with the matrix \(U\).
The two matrices \( U_1 \in \mathbb{R}^{ m \times m } \) and \( U_2 \in \mathbb{R}^{ n \times n } \) contain the parameters for reconstructing the original feature vectors:
Each row of \( U_1 \) is a speaker parameter and each row of \( U_2 \) a pose parameter.
Basically, each speaker parameter represents a point in the \(m\)-dimensional speaker subspace and each pose parameter a point in the \(n\)-dimensional pose subspace that are linked together by the tensor \(C\).
We remark that, compared to a \ac{PCA} model, such a multilinear model offers the advantage that it aims at capturing anatomical and articulation related shape variations separately.

The tensor \(C\) can be used to create new positional data for provided parameters \(\vec{s}\) and \(\vec{p}\):
\begin{equation}
  v(\vec{s}, \vec{p}) = \mu + C \times_1 \vec{s} \times_2 \vec{p}
\end{equation}
where \( \mu \) is a feature vector consisting of the positional data that corresponds to the mean mesh of the tongue shape collection.
This generated information can be utilized to construct a new tongue shape:
We reconstruct the vertex set by using the created positional data and combine it with the original face set to obtain our mesh.
More details on how the model was derived and evaluated can be found in \cite{Hewer2018CSL}.

\begin{figure}
  \centering
  \includegraphics[width=0.5\linewidth]{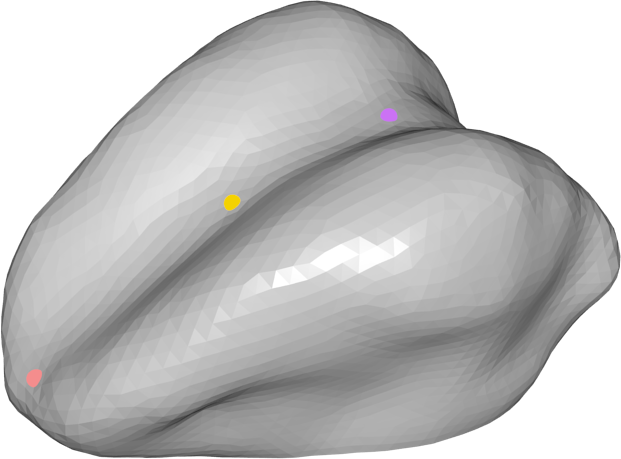}
  \caption{Rendered tongue model mesh, highlighting three vertices selected to correspond to the tongue coils in the \ac{EMA} data;
    pink: T1, yellow: T2, purple: T3 (cf.\ \cref{fig:ema-layout}).}
  \label{fig:tongue-mesh}
\end{figure}

In our framework, we use this model to register data of an \ac{EMA} corpus in order to obtain the corresponding parameters, which is done as follows:
In a first step, we manually align the \ac{EMA} data to the model space by using a provided reference coil.
As we want to register the \ac{EMA} data, we have to decide which coil corresponds to which vertex of the model mesh.
This process is done in a semi-supervised way:
The parameters are first set to random values and the associated mesh \( f ( \vec{s}, \vec{p} ) \) is generated.
Next, for each considered coil the nearest vertex on the mesh is found.
We then refine these correspondences iteratively by fitting the model to the coils and updating the nearest vertices.
In the end, we keep the correspondences that resulted in the smallest average Euclidean distance.
Finally, we inspect the result manually and repeat the experiment if the correspondences appear to be wrong.
The tongue model mesh is shown in \cref{fig:tongue-mesh}, highlighting the vertices selected to correspond with the three tongue coils in the \ac{EMA} data.
With these estimated correspondences, we fit the multilinear model to each considered \ac{EMA} data frame of the corpus by minimizing the energy:
\begin{equation}
  \label{eq:model-fitting}
  E(\vec{s}, \vec{p} ) = E_\text{Data} ( \vec{s}, \vec{p} ) + \alpha\ E_\text{SC} ( \vec{s} ) + \beta\ E_\text{PS} ( \vec{p} )
\end{equation}

The data term \(E_\text{Data} (\cdot) \) measures the distances between the selected vertices of the generated mesh \( f( \vec{s}, \vec{p} ) \) and the corresponding coil positions.
The speaker consistency term \( E_\text{SC} (\cdot) \) weighted by \( \alpha > 0 \) generates energy if the current speaker parameter differs from the one of the previous time step.
The remaining term, the pose smoothness term \( E_\text{PS} (\cdot) \) weighted by \( \beta > 0 \) fulfills a similar role:
It penalizes changes of the pose parameter over time.
As a minimizer of this energy is the best compromise between those mentioned assumptions, the fitting results will be close to the data and show smooth transitions over time.
The degree of smoothness can be controlled by adjusting the weights \( \alpha \) and \( \beta \).
As the multilinear model can be used to measure the probability of generated shapes, we can also choose how far the results are allowed to deviate from the model mean:
We limit the possible values for each entry of the parameters to an interval \( [ m_i - c\ \sigma_i, m_i + c\ \sigma_i ] \) where \( c > 0 \), \( m_i \) is the mean and \( \sigma_i \) the standard deviation of the corresponding entry in the training set of tongue meshes.
In order to obtain a minimizer, we use a quasi-Newton solver \cite{Liu1989} that supports limiting the solution to the given intervals.

\subsection{Multimodal Statistical Parametric Speech Synthesis}
\label{sec:hts}
\acreset{HTS}

The \ac{HTS} framework first presented by \citet{Zen2005} is a standard statistical parametric speech synthesis system.
The architecture comprises four main parts:
\begin{enumerate}[nosep]
\item the parametrization of the signal,
\item the training of the models,
\item the parameter generation, and
\item the signal rendering.
\end{enumerate}

The focus of our study impacts the parametrization (a) and the rendering (d) stages.
Therefore, we use the standard training stage (b), described in \cite{Zen2005}, and the standard parameter generation algorithms (c), described in \cite{Tokuda2000}.

The parametrization of the signal can be performed using any suitable signal processing tool, as long as it is kept consistent with the signal rendering.
In the standard procedure, this is generally accomplished by coupling STRAIGHT \cite{Kawahara1999} with a \ac{MLSA} filter \cite{Fukada1992}.
First, STRAIGHT is used to extract the spectral envelope, the \ac{F0}, and the aperiodicity.
Generally, the \ac{F0} values are transformed into the logarithmic domain, to be more consistent with human hearing.
Since the number of coefficients used of the spectral envelope and the aperiodicity is too high, the \ac{MLSA} filter is used to parametrize these coefficients and to obtain the \ac{MGC} and the \ac{BAP}, respectively.

\begin{figure}
  \centering
  \includegraphics[width=\linewidth]{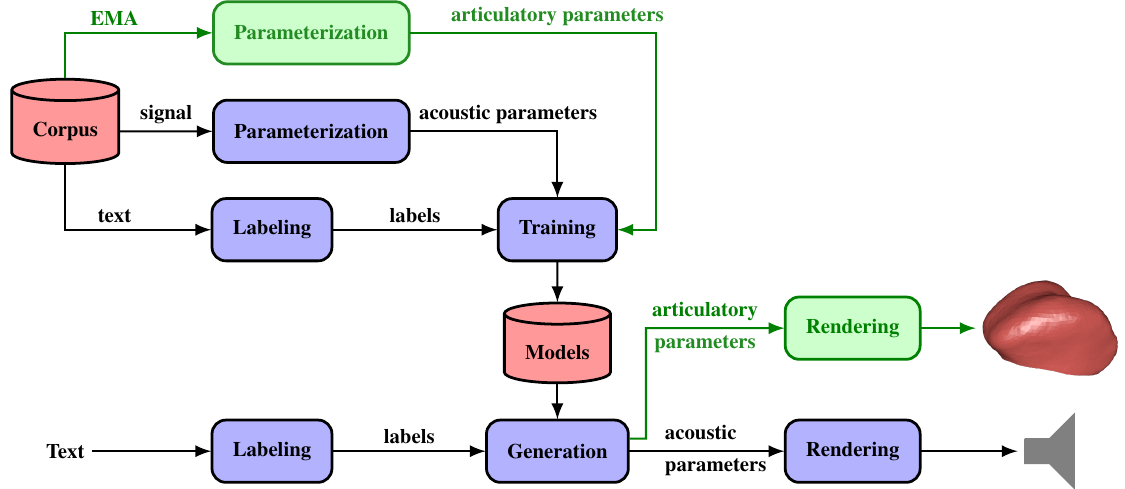}
  \caption{Architecture for multimodal \acs{HMM} based synthesis adapted from \cite{Zen2005};
    the multimodal extensions are highlighted.}
  \label{fig:arch-hts-m2}
\end{figure}

In this study, we propose to not only consider the parametrization of the acoustic signal but also the parametrization of speech articulation.
In previous studies \cite{Ling2009,Ling2010IS,Ling2010SpeCom}, \ac{EMA} data was used as the articulatory representation.
In the present study, we work towards replacing the \ac{EMA} data by the tongue model parameters.
Therefore, our goal is to train on the trajectories of the tongue model parameters using the standard \ac{HTS} framework as presented by \citet{Zen2005}.
The training models in \ac{HTS} are \acp{HMM}, at a phone level, whose observations are composed by decision trees.
The leaves of the decisions trees are \acp{GMM} which are used to produce the parameters at the generation level.
The generation level consists of applying the algorithm presented by \citet{Tokuda2000}.
\cref{fig:arch-hts-m2} presents the details of the modified architecture.

\section{Experiments}
\label{sec:experiments}

\subsection{Multilinear Model}

As the database for deriving the multilinear model, we used \ac{MRI} data from the Ultrax project \citep{Richmond2012} (11 speakers) and combined it with the data of \citet{Baker2011} (1 speaker), which was recorded as part of the Ultrax project, but released separately.
In the end, the resulting tongue mesh collection contained, for each speaker, estimated shapes for the phone set \textipa{[i, e, E, a, A, 2, O, o, u, 0, @, s, S]}.
Accordingly, the resulting multilinear model has \SIlist{12;13}{\ac{DoF}} for the anatomy and tongue pose, respectively.
The tongue mesh we used for the template matching was manually extracted from one \ac{MRI} scan, made symmetric to remove some bias towards the original speaker, and finally remeshed to be more isotropic.
It consists of \num{3100} vertices, \num{6102} faces, and has a spatial resolution of \SI{1.87}{\mm}.

\subsection{Database}
\label{sec:data}

\begin{figure}[t]
  \centering
  \includegraphics[width=0.7\linewidth]{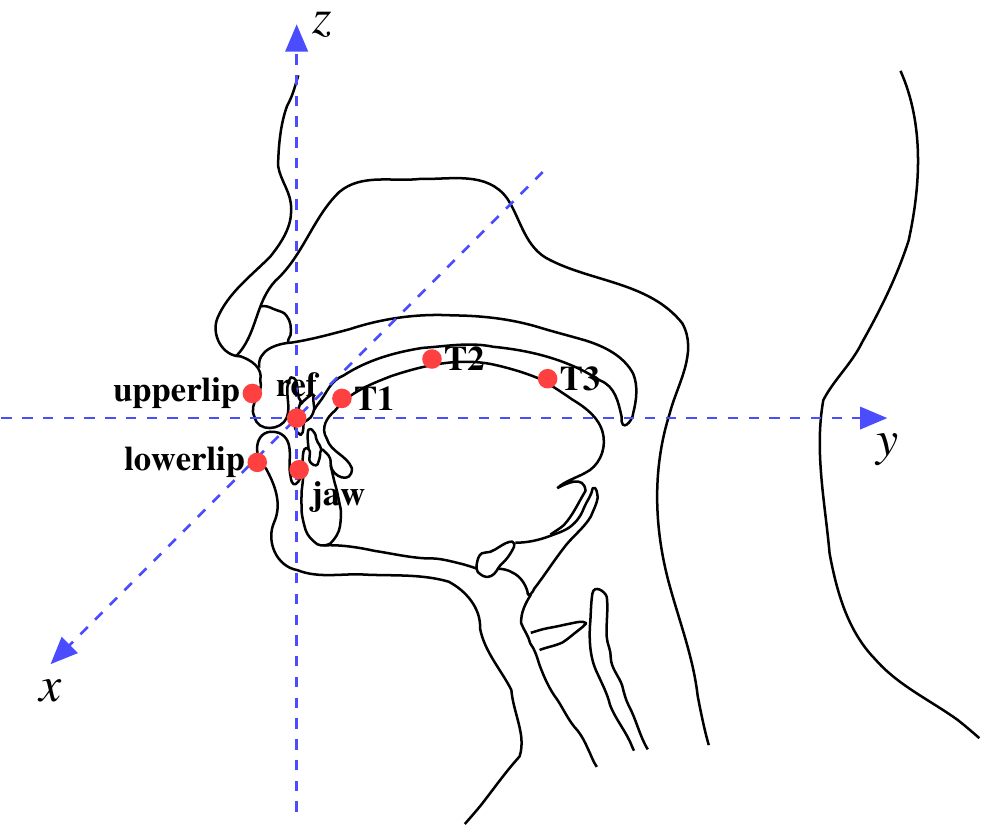}
  \caption{\acs{EMA} coil layout in the \enquote{day 1} subset of the \emph{mngu0} corpus.
    All coils are close to the mid-sagittal plane.
    The \textbf{ref} coil on the upper incisors forms the origin of the coordinate space;
    the horizontal and vertical axes represent the $y$ and $z$ dimensions in the data, respectively, while the $x$ axis is perpendicular to the image plane.
    Adapted from \cite{Richmond2011}.}
  \label{fig:ema-layout}
\end{figure}

The data used for the experiments in this study is taken from the \emph{mngu0} corpus, specifically the \enquote{day 1} \ac{EMA} subset \cite{Richmond2011}, which contains acoustic recordings, time-aligned phonetic transcriptions, and \ac{EMA} motion capture data (sampled at \SI{200}{\hertz} using a Carstens AG500 articulograph).%
\footnote{From the \emph{mngu0} website, \url{http://mngu0.org}, we downloaded the following distribution packages:
  \begin{enumerate}[nosep]
    \item Day1 basic audio data downsampled to \SI{16}{\kilo\hertz} (v1.1.0)
    \item Day1 basic \acs{EMA} data, head corrected and unnormalized (v1.1.0)
    \item Day1 transcriptions, Festival utterances and ESPS label files (v1.1.1)
  \end{enumerate}}
We selected the \enquote{basic} (as opposed to the \enquote{normalized}) release variant of the \ac{EMA} data, because it preserves the silent (i.e., non-speech) intervals, as well as the \ac{3D} nature and true spatial coordinates of the sensor data (after head motion compensation).
The \ac{EMA} coil layout for this data is shown in \cref{fig:ema-layout};
the coils are explained in \cref{tab:ema-coils}.

In order to manipulate the \ac{EMA} data more flexibly, the files were first converted from the binary \ac{EST} format to a JSON structure.
Invalid values (i.e., \texttt{NaN}) were replaced by linear interpolation.
No further modification, in particular no smoothing, was applied.

From the provided acoustic data, signal parameters were extracted using STRAIGHT \cite{Kawahara1999} with a frame rate of \SI{200}{\hertz}, matching that of the \ac{EMA} data.
As we follow the standard \ac{HTS} methodology, we also kept the same parameters.
Therefore, our signal parameters are \SI{50}{\ac{MGC}}, \SI{25}{\ac{BAP}}, and one coefficient for the \ac{F0}.

\begin{figure}
    \centering
    \includegraphics[width=\linewidth]{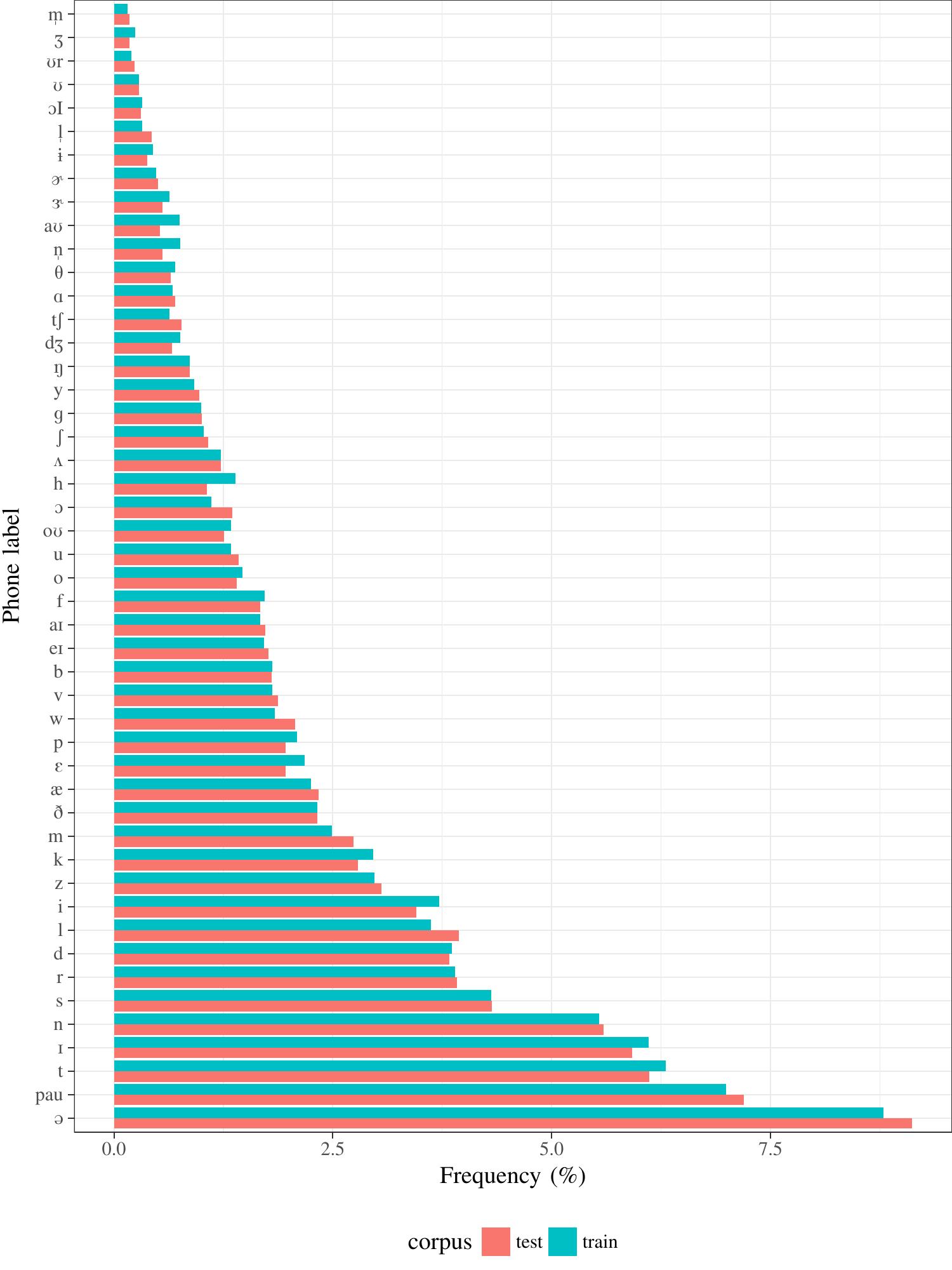}
    \caption{Distribution of phones across the training and test sets.}
    \label{fig:dist_corpus}
\end{figure}

From the \num{1354} utterances in the data, \num{152} (\SI{11.2}{\percent}, around \SI{10}{\minute}) were randomly selected and held back as a test set;
the remaining \num{1202} utterances (around \SI{105}{\minute}) were used as the training set to build \ac{HTS} synthesis voices.
A comparison of phone distributions in the training and test sets shows a satisfactory match (cf.\ \cref{fig:dist_corpus}).

\begin{table}[t]
  \centering
  \caption{\acs{EMA} Coil Labels and Locations in the \enquote{Day 1} Subset \\
    of the \emph{mngu0} Corpus.}
  \smaller
  \begin{tabulary}{\linewidth}{LC}
    \toprule
    \textbf{Label} & \textbf{Location} \\
    \midrule
    \textbf{T1} & Tongue tip \\
    \textbf{T2} & Tongue body \\
    \textbf{T3} & Tongue dorsum \\
    \textbf{upperlip} & Upper lip \\
    \textbf{lowerlip} & Lower lip \\
    \textbf{ref} & Upper incisor \\
    \textbf{jaw} & Lower incisor \\
    \bottomrule
  \end{tabulary}
  \label{tab:ema-coils}
\end{table}

\subsection{Acoustic Synthesis}
\label{sec:straight}

\begin{table}
    \centering
    \caption{Global Evaluation Measures for the Acoustic Synthesis Baseline Conditions.}
    \includegraphics{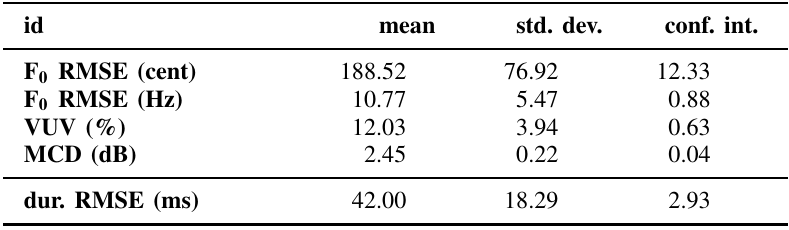}
    \label{tab:straight}
\end{table}

As a baseline, we first built a conventional \ac{TTS} system using the acoustic data only.
This served mainly to validate our voicebuilding process and ensure that the transcriptions provided, and labels generated from them, along with the acoustic signal parameters, were able to generate audio of sufficient quality.
Accordingly, we did not undertake a formal subjective listening test, and instead evaluated this baseline experiment using objective measures only.

We synthesized the \num{152} utterances in the test set using two conditions.
The first condition is the standard synthesis process.
This condition allows us to evaluate the duration accuracy.
For the second condition, we imposed the acoustic phone durations from the provided transcriptions to allow direct comparison with the natural recordings.
For the following experiments, we synthesized both conditions as well.
The objective evaluation was conducted based on the following metrics.

For the duration evaluation, we calculated the duration \ac{RMSE} at the phone level (in \si{\ms}) between the reference duration and the one synthesized using the first condition.

Considering the other coefficients, we compared the synthesis result ($s$), achieved using the second condition, to the reference ($r$) present in the test corpus.
As the duration was imposed, we have the same number $T$ of frames for the produced utterance and the reference one.
To evaluate the \ac{F0}, we used three measures:
the \ac{VUV} error rate percentage $VUV(r,s)$ \cref{eq:vuv} to check the prediction of the \ac{F0}, the \ac{RMSE} in \si{\hertz} \cref{eq:rmshz}, and the \ac{RMSE} in cent \cref{eq:rmscent}.
The latter measure focuses on the frames which are voiced in both conditions (original and predicted \ac{F0}).
Furthermore, it is a log scale measure adapted to the human perception.
\begin{equation}
    v(x, y) = \left\{
      \begin{array}{ll}
        0 & \text{x \& y are both voiced/unvoiced}\\
        1 & \text{otherwise} \\
      \end{array}
    \right.
\end{equation}
\begin{equation}
  \label{eq:vuv}
    VUV(r, s) = (\sum_{t=1}^T v(r_t, s_t) / T) * 100
\end{equation}
\begin{equation}
  \label{eq:rmshz}
    RMSE_{\si{\hertz}}(r, s) = \sqrt{\sum_{t=1}^T (r_t - s_t)^2 / T}
\end{equation}
\begin{equation}
  \label{eq:rmscent}
    RMSE_{\text{cent}}(r, s) = \sqrt{ 1200 * \sum_{t=1}^T (log(r_t) - log(s_t))^2 / T}
\end{equation}
Finally, to evaluate the spectral envelope production, we computed the \ac{MCD} between the \ac{MGC} vectors of dimension $M$ in \si{\decibel}:
\begin{equation}
    d(x, y) = \sum_{m=2}^M(x(m) - y(m))^2
\end{equation}
\begin{equation}
  \label{eq:mcd}
      MCD(r, s) = \frac{10}{\ln 10} * \sqrt{2} * \sqrt{\sum_{t=1}^T d(r_t, s_t) / T}
\end{equation}

Except for the duration, all parameters were evaluated at the frame level.
Based on these measures, we can compare our results to previous studies, such as the one presented by \citet{Yokomizo2010}.

The results of this evaluation are given in \cref{tab:straight} and comprise the mean, standard deviation, and confidence interval with a $p$ value at \SI{5}{\percent}.
Compared to \cite{Yokomizo2010}, we achieved slightly better results, notwithstanding the different dataset.
Therefore, we can conclude that our acoustic prediction is consistent with the state of the art in \ac{HTS}.

\subsection{Combined Acoustic and \ac{EMA} Synthesis}
\label{sec:straight_ema}

Adopting the paradigm of early multimodal fusion, we combined the acoustic signal parameters with the \ac{3D} positions of the seven \ac{EMA} coils shown in \cref{tab:ema-coils}, increasing the vector size by \num{21}, to \num{97} parameters per frame.
Using the \ac{HTS} framework, we then built another \ac{TTS} system from this multimodal data.

Synthesizing the test set in this way, we obtained, in addition to the audio, synthetic trajectories of predicted \ac{EMA} coil positions.
To evaluate the combined acoustic and \ac{EMA} synthesis, we computed the same objective measures as in \cref{sec:straight}.
We also computed the Euclidean distance in space between the observed and predicted positions for the \ac{EMA} coils.
Finally we computed the \ac{RMSE} between the dynamics of the trajectories of the coils using a unit of millimeters per frame (\si{\mm\per frame}).
The results of this evaluation are given in \cref{tab:straight_ema}.
We see that the differences in the acoustic measures compared to the acoustic-only synthesis (cf.\ \cref{tab:straight}) are negligible.

\begin{figure*}
  \centering
  \includegraphics[width=0.8\linewidth]{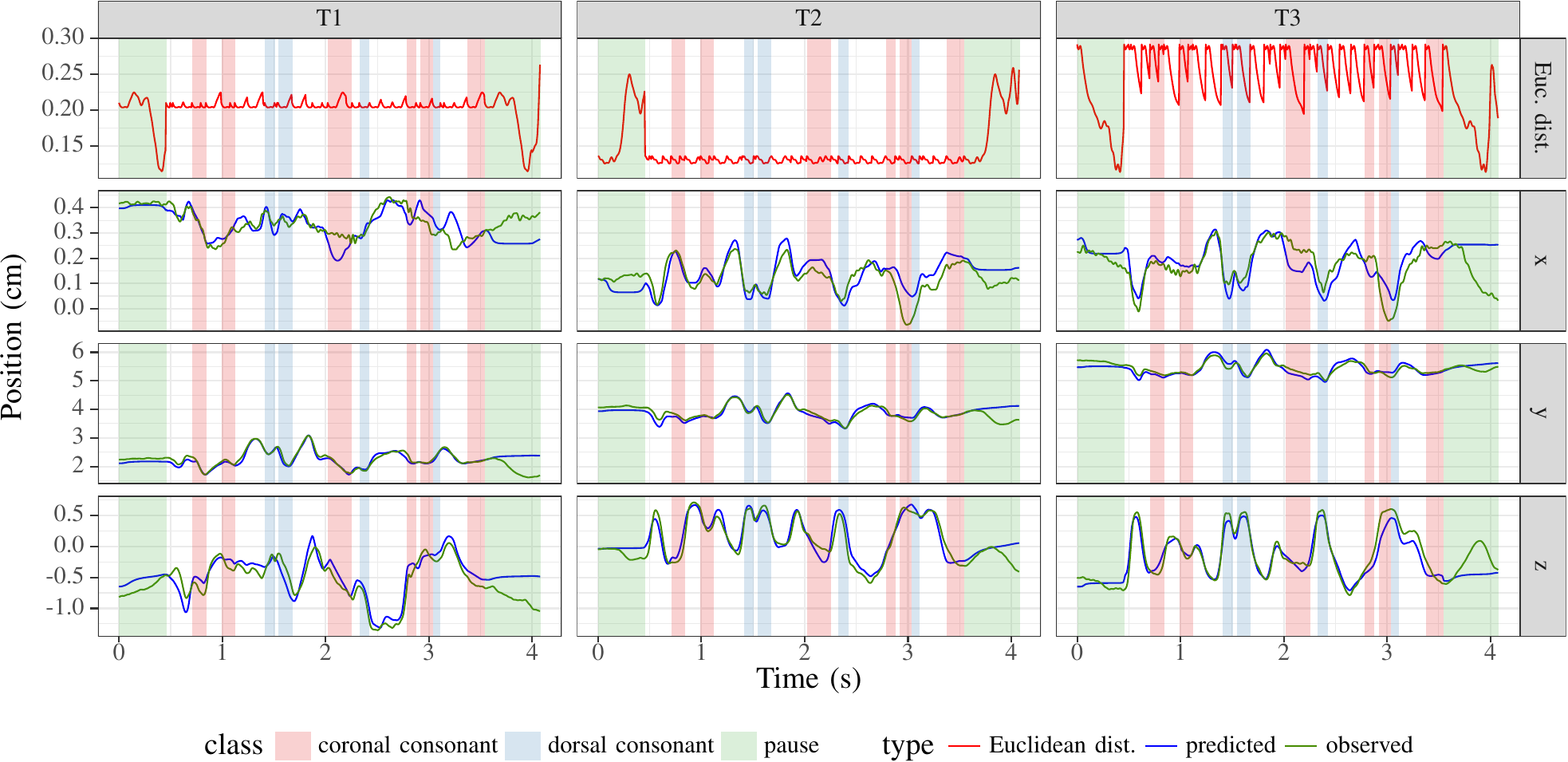}
  \caption{Observed and predicted position trajectories (along the $x$, $y$, and $z$ axis), and Euclidean distance (top), for the tongue \ac{EMA} coils (T1, T2, T3) for one test utterance, using combined acoustic and \acs{EMA} synthesis.
    The utterance is \enquote{Because these deer are gregarious, they go about in groups}.
    Based on the provided transcriptions, intervals containing silent (pause) and coronal and dorsal consonants have been highlighted.}
  \label{fig:example-straight_ema}
\end{figure*}

The comparison between the observed and predicted trajectories for one test utterance is illustrated in \cref{fig:example-straight_ema}.
The observed and predicted (synthesized) positions of the three tongue coils are shown in each of the three dimensions in the data, along with the Euclidean distance.
Silent intervals and consonants classified as coronal \textipa{[t, d, n, l, s, z, S, Z, T, D]} and dorsal \textipa{[g, k, N]}, based on the provided phonetic transcription, have been highlighted.
This helps visualize the correspondence between gestures of the tongue tip (coil T1) and tongue back (coils T2 and T3) for coronal and dorsal consonants, respectively, and the phonetic units they produce.

Several points merit discussion.
First of all, there are large mismatches between the observed and predicted tongue \ac{EMA} coil positions during the silent (pause) intervals at the beginning and end of the utterance.
This can be attributed to the fact that the wide range of the speaker's tongue movements during non-speech intervals are not distinguished in the provided annotations, but invariably labeled with the same pause symbol.
However, there are at least two very distinct shapes for the tongue during such silent intervals, including a \enquote{rest} and a \enquote{ready} position (just before speech is produced), in addition to other complex movements such as swallowing.
In the absence of distinct labels corresponding to these positions and movements, none of this silent variation can be captured by the \acp{HMM} trained on this data;
instead, the tongue coils are unsurprisingly predicted to hover around global means.

Secondly, there is noticeable oversmoothing and target extrema are not always quite reached.
This can typically be attributed to the \ac{HMM} based synthesis technique, despite the integration of global variance.
The dynamics, however, are well represented, and the predicted positional trajectories, as well as their derivatives, match the observed reference quite closely.

The $x$ axis appears to suffer from a greater amount of prediction error than the $y$ or $z$ axes.
However, it should be noted that the positional variation along the $x$ axis is an order of magnitude smaller than that along the $y$ axis.
It must also be borne in mind that nearly all of the speech-related movements occur in the mid-sagittal plane, represented by the $y$ (anterior/posterior) and $z$ (inferior/superior) axes;
variation along the $x$ axis corresponds to lateral movements, which are infrequent during speech.%
\footnote{Incidentally, the \enquote{normalized} release variant of the \emph{mngu0} \ac{EMA} dataset follows this rationale and consists of flattened, \ac{2D} data, with all coil positions projected onto the mid-sagittal plane.}
Having said that, the $x$ axis can serve to illustrate the physical coil locations on the tongue in the \enquote{day 1} recording session;
to wit, the tongue tip coil is actually attached out of plane, a few millimeters to one side.

The Euclidean distances \emph{during} speech are in the millimeter range, indicating that the predictions of \ac{EMA} coil positions are accurate to within the precision of the \ac{EMA} measurements themselves.
However, there appears to be a certain amount of fluctuation with a more or less regular range and shape.
The peaks of this fluctuation appear to correlate with spikes in the rms channels of the provided \ac{EMA} data, which supports the hypotheses that it is either an artifact of the algorithm which calculates the coil positions and orientations from the raw amplitudes \citep{Stella2012}, or measurement noise in the articulograph itself \citep{Kroos2012}, or, conceivably, a combination of both factors.
Of course, the noise in the Euclidean distance analysis is a direct consequence of our decision to refrain from smoothing the provided \ac{EMA} data.%
\footnote{Perhaps the rms jitter in the unsmoothed measurements could also be exploited in adaptive \ac{EMA} denoising.}

\begin{table}
    \centering
    \caption{Global Evaluation for the Combined Acoustic and \acs{EMA} Synthesis.}
    \includegraphics{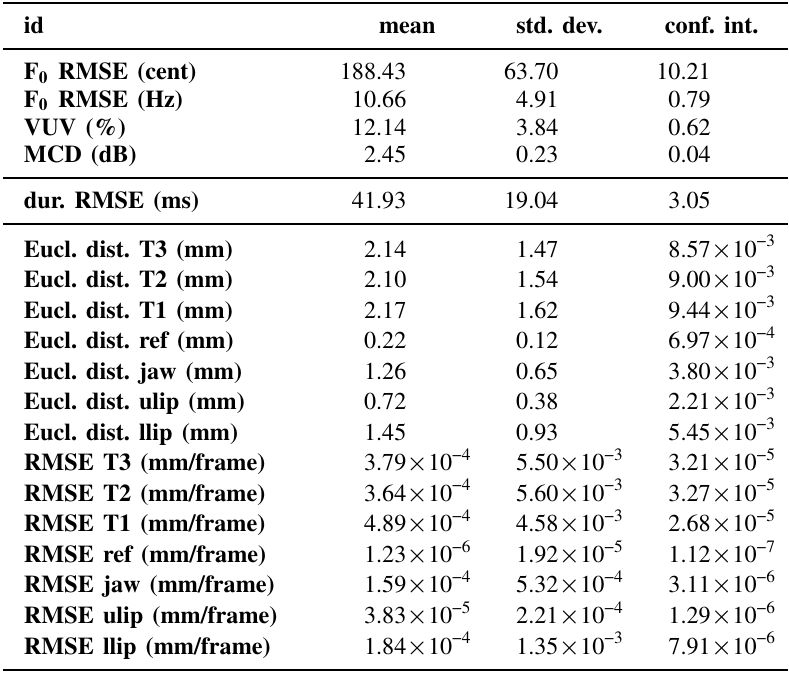}
    \label{tab:straight_ema}
\end{table}

\subsection{\ac{EMA} Synthesis}
\label{sec:ema}

\begin{figure*}
  \centering
  \includegraphics[width=0.8\linewidth]{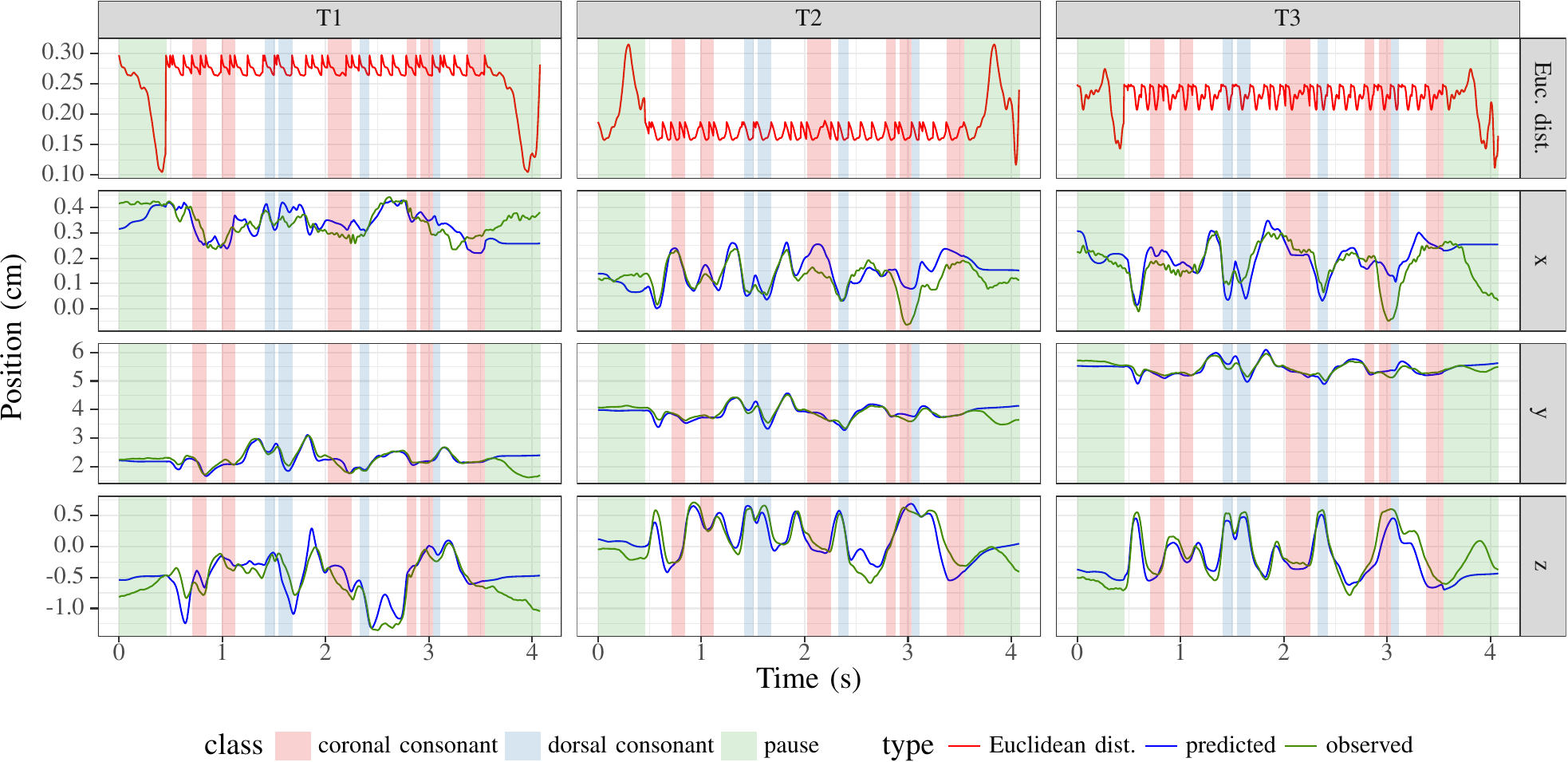}
  \caption{One test utterance produced using \acs{EMA}-only synthesis;
    all other details are the same as in \cref{fig:example-straight_ema}.}
  \label{fig:example-ema}
\end{figure*}

\begin{table}
    \centering
    \caption{Global Evaluation for the \acs{EMA}-Only Synthesis.}
    \includegraphics{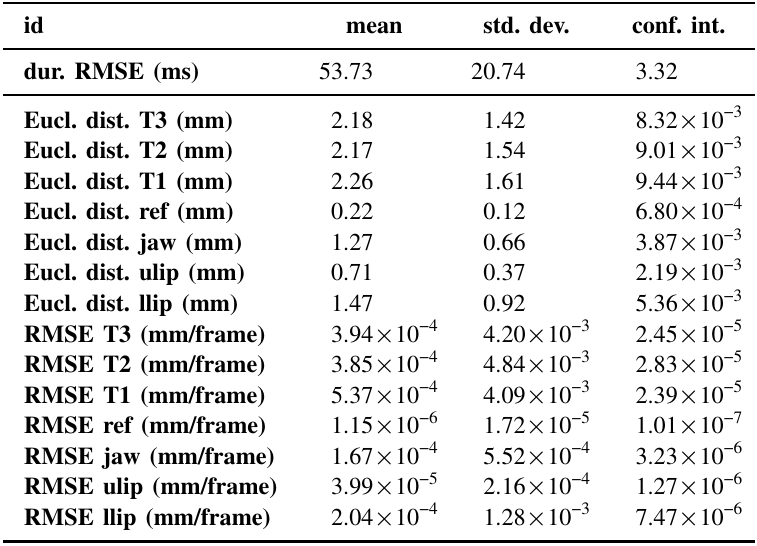}
    \label{tab:ema}
\end{table}

While the combined acoustic and \ac{EMA} synthesis produced satisfactory results, the requirement to train the system on a multimodal dataset such as \emph{mngu0} represents a significant drawback;
compared to the reasonably wide availability of conventional, acoustic databases designed for speech synthesis, the number of suitable articulatory databases is extremely low.
Encouraged by the practical equivalence in the evaluation of the acoustic measures described in \cref{sec:straight,sec:straight_ema}, we therefore considered the question of decoupling the \ac{EMA} synthesis completely from the acoustic data.
Accordingly, we used the \ac{HTS} framework to build another \ac{TTS} system trained only on the \ac{EMA} data, without the acoustic parameters.

Under this condition, the evaluation of the duration \ac{RMSE} and Euclidean distances between the predicted and observed \ac{EMA} coils, computed using the formula given by \cref{eq:rmshz}, is given in \cref{tab:ema}.
As we can see, the results are nearly identical to those in \cref{tab:straight_ema}, which confirms the validity of this approach.
\cref{fig:example-ema} visualizes the comparison between the observed and predicted trajectories for one test utterance.

\subsection{Tongue-only \ac{EMA} Synthesis}
\label{sec:ema_tongue}

\begin{figure*}
  \centering
  \includegraphics[width=0.8\linewidth]{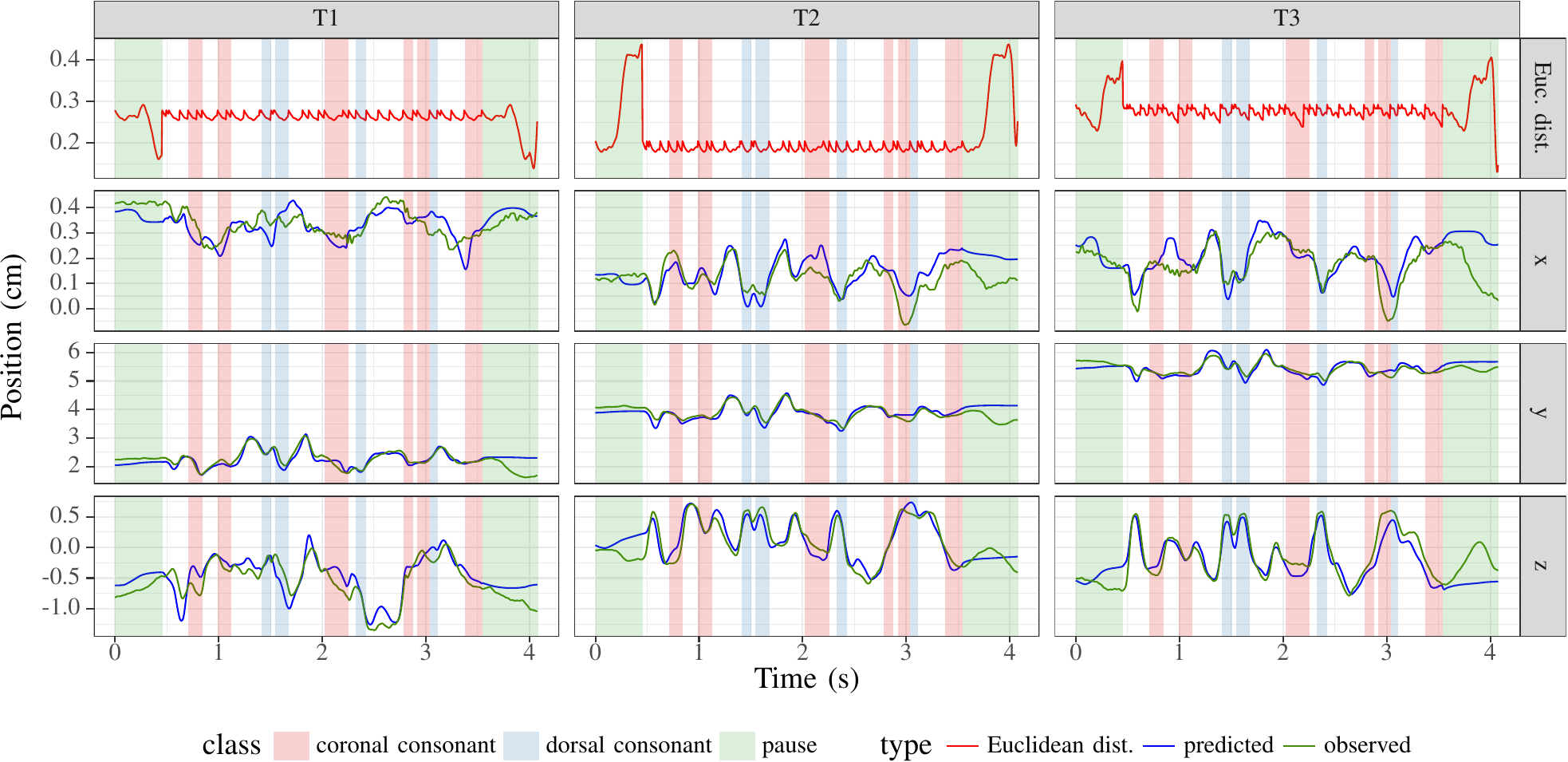}
  \caption{One test utterance produced using \acs{EMA}-only synthesis restricted to the tongue coils;
    all other details are the same as in \cref{fig:example-straight_ema}.}
  \label{fig:example-ema_tongue}
\end{figure*}

\begin{table}
    \centering
    \caption{Global Evaluation for the \acs{EMA}-Only Synthesis Restricted to the Tongue Coils.}
    \includegraphics{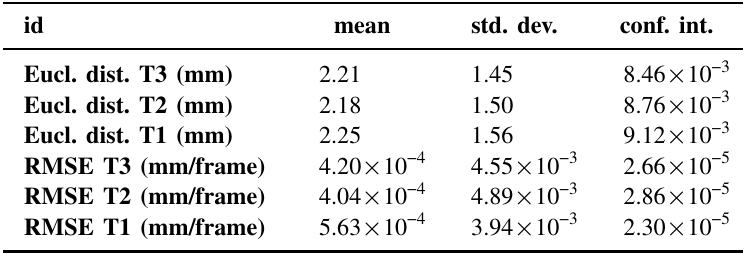}
    \label{tab:ema_tongue}
\end{table}

In order to focus on the tongue in the following section, we first needed to investigate how far the tongue coil \ac{EMA} positions can be predicted in isolation from the remaining \ac{EMA} coils.
To this end, we created a modified version of the \ac{TTS} system described in the previous section, by including \emph{only} the tongue coils (T1, T2, and T3), and excluding the rest of the \ac{EMA} data from the training set.

\Cref{tab:ema_tongue} gives the evaluation of the \ac{EMA} synthesis restricted to the three tongue coils.
Comparing these results with those in \cref{tab:ema}, we observe that the values are virtually identical, which confirms the validity of this approach.
As before, the comparison between the observed and predicted trajectories for one test utterance is shown in \cref{fig:example-ema_tongue}.
It should be noted that despite the removal of the \ac{EMA} coil on the lower incisor, some residual jaw motion is implicitly retained in the movements of the tongue coils.

\subsection{Model-based Tongue Motion Synthesis}
\label{sec:weights}


\begin{figure*}
  \centering
  \includegraphics[width=0.8\linewidth]{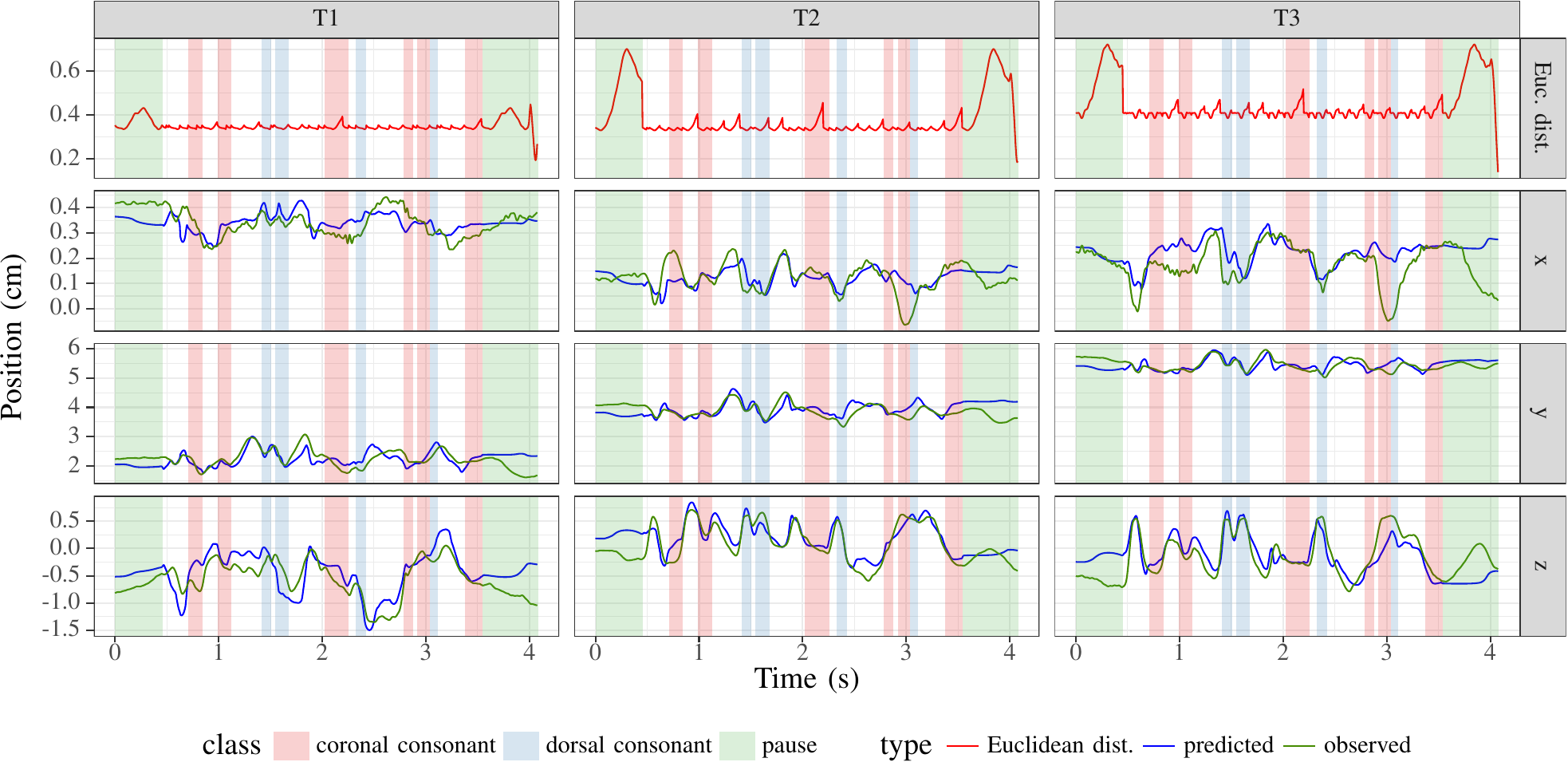}
  \caption{One test utterance produced using the tongue model parameters synthesis;
    all other details are the same as in \cref{fig:example-straight_ema}.}
  \label{fig:example-weights}
\end{figure*}

\begin{table}
    \centering
    \caption{Global Evaluation for the Tongue Model Parameters Synthesis.}
    \includegraphics{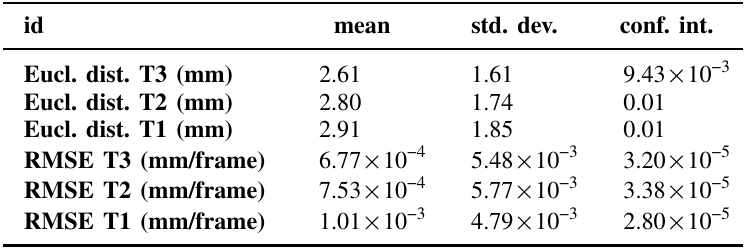}
    \label{tab:weights}
\end{table}

\begin{figure*}
  \centering
  \includegraphics[width=0.8\linewidth]{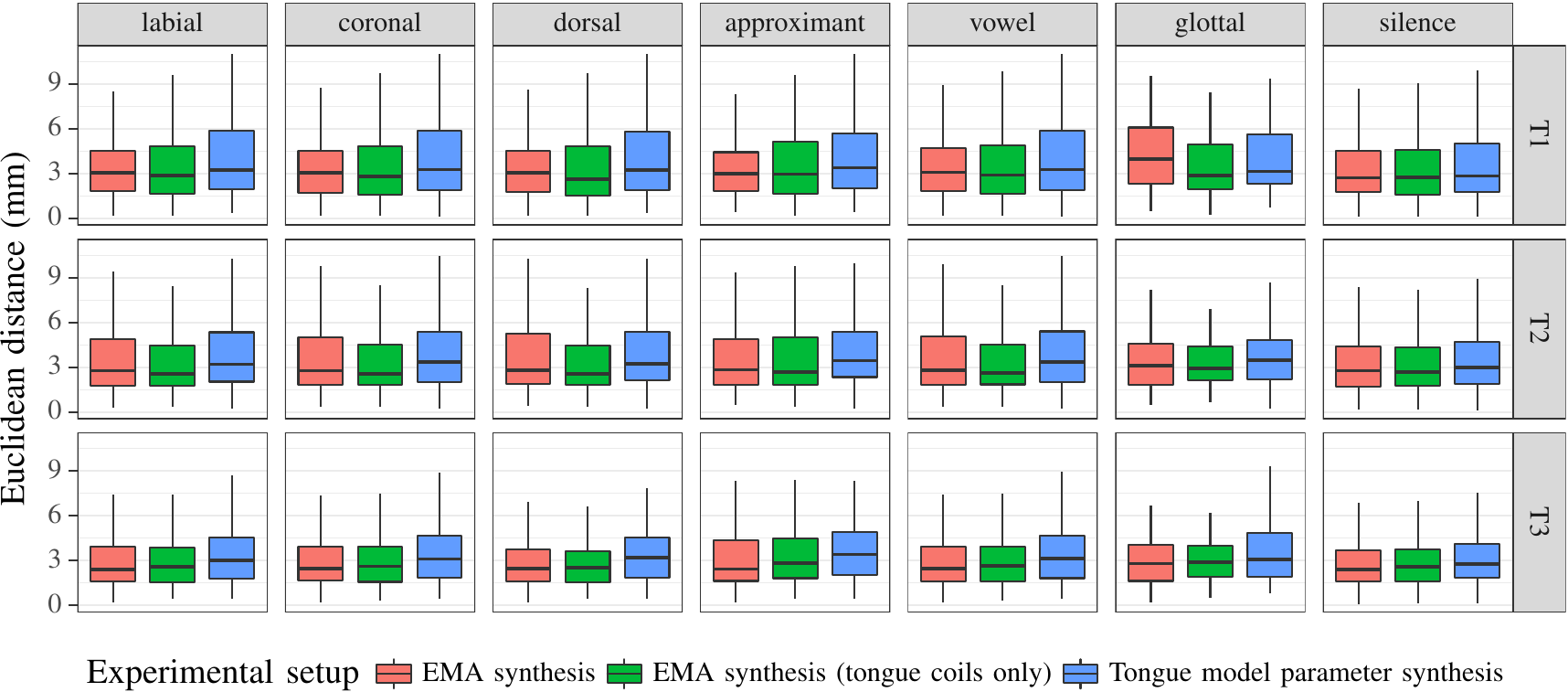}
  \caption{Distributions of Euclidean distances between observed and predicted tongue \ac{EMA} coil positions for each experimental \acs{TTS} setup, split by phone class and tongue \acs{EMA} coil.}
  \label{fig:phone-class-distributions}
\end{figure*}

At last, having verified that the \ac{HTS} framework can be used to synthesize audio and predict the movements of three tongue \ac{EMA} coils using \emph{separate} models trained on the \emph{mngu0} database, we prepared a new kind of \ac{TTS} system to predict the shape and motion of the entire tongue surface, by integrating the multilinear model into the process.

To this end, we first estimated the anatomical features \( \vec{s} \) (cf.\ \cref{sec:tonguemodel}) of the speaker in the \emph{mngu0} dataset as follows:
We used the upper incisor coil as a reference and estimated the correspondences between the three tongue coils and the model vertices, chosen as described in \cref{sec:tonguemodel}.
During this correspondence optimization, we used \(c = 0.25\).
Thus, we limit the admissible values for each entry of the model parameters to the interval \( [ m_i - 0.25\ \sigma_i, m_i + 0.25\ \sigma_i ] \) where \( m_i \) is the mean and \( \sigma_i \) the standard deviation of the corresponding model parameter.
By using such a small interval, we try to prevent overfitting during this step.
Afterwards, we fitted the model to all \ac{EMA} data frames and stored the obtained parameter values.
Here, we used the speaker consistency weight \( \alpha = 20 \) and the pose smoothness weight \( \beta = 10 \) in the fitting energy.
Thus, we demanded very smooth transitions in this case and especially penalized changes of the speaker's anatomy over time.
In this step, we used \( c = 3 \) to give the approach some freedom during the fitting.
We then averaged all obtained speaker parameters to get an estimate of the considered speaker's anatomical features.

Next, we again fitted the model to all \ac{EMA} data frames where, this time, we fixed the speaker parameter \(\vec{s}\) to the estimated anatomy.
We note that this approach causes the multilinear model to behave like a single-speaker \ac{PCA} model.
This time, we used the weight \( \beta = 1 \) to increase the influence of the data term.
However, we decided to use \( c = 2 \) this time to motivate the approach to consider more plausible shapes.

We note that the settings for the fitting were selected manually by an expert.
Of course, this selection might be optimized for the used \ac{EMA} dataset by performing a thorough analysis.

The pose parameters resulting from this fitting step were taken as the training data, and we used the \ac{HTS} framework to build a new \ac{TTS} system that predicts the tongue model parameter values directly from the input text.

To evaluate the performance of this system against the reference \ac{EMA} data, we extracted the spatial coordinates of the vertices assigned during the adaptation step (see above) to produce synthetic trajectories that served as a virtual surrogate for predicted \ac{EMA} data.

We evaluated this synthetic \ac{EMA} data against the reference as before;
\Cref{tab:weights} provides the Euclidean distances between the predicted and observed \ac{EMA} coils, and one test utterance is visualized in \cref{fig:example-weights}.
It should be noted that the tongue model itself contains a temporal smoothing term, which ensures that a noisy sequence of input frames does not cause the \ac{3D} mesh to change shape or position too rapidly;
however, this extra smoothing contributes to widespread target undershoot in the comparison.
Overall, the results of this evaluation are very promising, and we can confirm that as far as possible, with only three surface points on the tongue, the animation of the full tongue appears to closely match the observed reference.

Finally, in order to compare the three experimental \ac{TTS} systems (trained without acoustic data), we analyzed the distribution of Euclidean distances between each system and the observed reference data over the entire test set;
the results are shown in \cref{fig:phone-class-distributions}.
The distances are slightly greater when the non-tongue \ac{EMA} coils are excluded, and greater still when the \ac{EMA} prediction is replaced by the direct synthesis of tongue model parameters.
However, overall, the distances remain in the same range, which indicates that the latter approach perform no worse than synthesis of \ac{EMA} data -- while adding the full \ac{3D} tongue surface into the synthesis process.

\section{Conclusion}
\label{sec:conclusion}

In this study, we have presented a new process of synthesizing acoustic speech and synchronized animation of a full \ac{3D} surface model of the tongue.
We used the \ac{HTS} framework with a single-speaker, multimodal articulatory database containing \ac{EMA} motion capture data.
First, we demonstrated a conventional, fused multimodal approach, then separated the two modalities while ensuring that the objective evaluation measures remained comparable.
Finally, we adapted a multi-linear statistical model of the tongue and integrated it into the \ac{TTS} process, and evaluated its accuracy by comparing the spatial coordinates of vertices on the model surface to the reference \ac{EMA} data from the original speaker's tongue movements.
The results are very encouraging, and we believe that this will enable multimodal \ac{TTS} applications that provide tongue animation with human-like performance.

It should be noted that the acoustic synthesis and predicted phone durations need not come from the same corpus as the one used for training the tongue model parameter synthesis system.
Under certain conditions, it would be straightforward to use a different, conventional \ac{TTS} system with speech recordings from a different speaker in combination with this tongue model parameter synthesis, perhaps adapting it in the speaker subspace automatically or by hand, to generate a multimodal \ac{TTS} application with plausible, speech-synchronized tongue motion, without the requirement of having articulatory data available for the target speaker.
In this way, it is possible to first synthesize the acoustic speech signal, and to provide the predicted acoustic durations to guide the synthesis of corresponding tongue model parameters, which are then used to render the animation of the \ac{3D} tongue model in real time.

However, there is clearly more work to be done, and in future research, we intend to refine and improve our system, and to evaluate it using human subjects who will rate it perceptually.
Such a study can include intelligibility, such as the contribution of visible tongue movements during degraded, noisy, or absent audible speech.
However, we also plan to assess the impact on perceived naturalness by integrating the tongue model into a realistic talking avatar (e.g., \citep{Taylor2012, Schabus2014}), and investigating the importance of naturalistic tongue movements for the overall impression of such avatars in multimodal spoken interaction scenarios with artificial characters.
This may also lead us to model distinct non-speech poses for the tongue, such as separate \enquote{rest} and \enquote{ready} positions.

Regarding the tongue model integration, we plan to further investigate such factors as the impact of reducing the dimensionality of the model subspaces on synthesis performance, optimizing the vertex correspondence with \ac{EMA} data, improving the fitting results by adjusting the weights for the smoothness terms, and exploring speaker adaptation using volumetric data, such as the \ac{MRI} subset of the \emph{mngu0} corpus \citep{Steiner2012}.

\newpage
\section*{Acknowledgments}

This work was funded by the German Research Foundation (DFG) under grants EXC~284 and SFB~1102.
The authors would like to thank Korin Richmond, Phil Hoole, and Simon King for creating and releasing the \emph{mngu0} database.
Studies such as the one described in this paper would not be possible without such high-quality, open databases.


\renewcommand{\bibfont}{\smaller}
\printbibliography


\vfill
\todototoc
\listoftodos

\end{document}